\RequirePackage{fix-cm}
\documentclass[twocolumn,epjc3]{svjour3}  
\smartqed  
\RequirePackage{graphicx}
\RequirePackage{subfigure, amssymb, amsmath}
\journalname{Eur. Phys. J. C}
\begin{document}

\title{Galactic Escape Speeds in Mirror and Cold Dark Matter Models
}

\titlerunning{Mirror and Cold Dark Matter}        

\author{Alvin J. K. Chua\thanksref{addr1}
       \and
       D. T. Wickramasinghe\thanksref{addr1} 
       \and
       Lilia Ferrario\thanksref{e3,addr1}
}

\thankstext{e3}{e-mail: Lilia.Ferrario@anu.edu.au}


\institute{Mathematical Sciences Institute, The Australian National University, ACT 0200, Australia \label{addr1} }

\date{Received: date / Accepted: date}

\maketitle

\begin{abstract}
The mirror dark matter (MDM) model of Berezhiani et al. has been shown
to reproduce observed galactic rotational curves for a variety of
spiral galaxies, and has been presented as an alternative to cold dark
matter (CDM) models. We investigate possible additional tests
involving the properties of stellar orbits, which may be used to
discriminate between the two models.  We demonstrate that in MDM and
CDM models fitted equally well to a galactic rotational curve, one
generally expects predictable differences in escape speeds from the
disc.

The recent radial velocity (RAVE) survey of the Milky Way has pinned
down the escape speed from the solar neighbourhood to $v_{\rm
  esc}=544^{+64}_{-46}$ km s$^{-1}$, placing an additional constraint
on dark matter models. We have constructed an MDM model for the Milky
Way based on its rotational curve, and find an escape speed that is
just consistent with the observed value given the current errors,
which lends credence to the viability of the MDM model.  The Gaia-ESO
spectroscopic survey is expected to lead to an even more precise
estimate of the escape speed that will further constrain dark matter
models. However, the largest differences in stellar escape speeds
between both models are predicted for dark matter dominated dwarf
galaxies such as DDO~154, and kinematical studies of such galaxies
could prove key in establishing, or abolishing, the validity of the
MDM model.

\end{abstract}

\section{Introduction}\label{intro}
Over the last decade, mirror matter has emerged as a promising
candidate for dark matter. The idea that there might be a hidden
``mirror sector'' of the Universe dates back to \cite{Lee1956}, was
subsequently expanded upon by \cite{Kobzarev1966}, and owes its modern
form (in the context of gauge theories) to \cite{Foot1991}. It is
based on the observation that parity symmetry is violated by the weak
interaction in the ``ordinary'' sector we observe, but may be restored
by the existence of a duplicate sector with the same fundamental
particles and microphysics (apart from the opposite handedness of the
weak interaction). Support for the existence of a mirror world may
also come from the recent experiments of \cite{Berezhiani2012} on
neutron oscillations.

Mirror dark matter (MDM) models linked to superstring inspired
unifications of the four forces in the early universe have been
discussed by \cite{Kolb1985}. The ordinary and mirror sectors are
coupled through gravity but not the other fundamental interactions,
which explains the intrinsic suitability of mirror matter as a dark
matter candidate.  Much research has been done on cosmological
signatures of mirror matter; these turn up in the context of early
Universe thermodynamics, Big Bang nucleosynthesis, primordial
structure formation/evolution, cosmic microwave background (CMB)
radiation and large-scale structure power spectra. Detailed reviews of
this work have been provided by \cite{Berezhiani2004} and
\cite{Ciarcelluti2010}.

Astrophysical signatures of mirror matter have also been looked at in
the context of galactic rotational curves, using a specific model for
mirror gravity developed by \cite{Rossi2008}, \cite{Berezhiani2009}
and \cite{Berezhiani2010} (henceforth BPR).  In this model, the
ordinary and mirror sectors are assumed to have separate metric
tensors (which are coupled to a third metric tensor), leading to a
Yukawa-like modification of the gravitational interaction between both
sectors. An ordinary matter test particle located at a radial distance
$r$ from a point-like source of ordinary mass $M_{1}$ and mirror mass
$M_{2}$ will feel a Yukawa-like potential
\begin{equation}\label{pot_eq}
\phi\left(r\right)=-\frac{G}{2r}\left(M_{1}+M_{2}+\left(M_{1}-M_{2}\right)e^{-\frac{r}{r_{m}}}\right),\end{equation}
where \emph{G} is the Newtonian gravitational constant and $r_{m}$ is
the Yukawa radius. The free-fall acceleration of the particle follows
as
\begin{equation}\label{accel}
g\left(r\right)=\frac{G}{2r^{2}}\left[M_{1}+M_{2}+\left(M_{1}-M_{2}\right)\left(1+\frac{r}{r_{m}}\right)e^{-\frac{r}{r_{m}}}\right].
\end{equation}

Although mirror matter is unable to form extended halos around
galaxies due to its collisional and dissipative nature, the combined
model of mirror matter and mirror gravity has been successfully fitted
to the observed rotational curves of several disc galaxies by BPR, on
the assumption that ordinary matter has a similar density profile to
its mirror counterpart. This makes it the most viable theory of MDM
that is presently available.

In this paper, we investigate some implications of the BPR model over
and above its demonstrated success in modelling rotation curves, and
suggest astrophysical signatures that may be observable and
potentially used to discriminate between the MDM and CDM models. In
Section 2, we use point-like source models to investigate the orbits
of ordinary matter particles in the presence of mirror matter, and
show that significant differences are expected in the nature of orbits
near the Yukawa radius. We then construct in Section 3 CDM models for
the galaxies NGC 2403, DDO 170 and DDO 154 (whose rotational curves
have previously been modelled using MDM), and calculate and compare
escape speeds from the stellar disc as a function of galactic radius
for both types of model. We show that the MDM models predict specific
values for the escape speeds that are generally different from those
in the CDM models. In Section 4, we construct MDM and CDM models for
the Milky Way, and find that the escape speed from the solar
neighbourhood in the MDM model is in good agreement with the current
observed value. We discuss and present our main conclusions in Section
5.
\section{Point-like source orbits}\label{point_s}

\begin{figure*}
\begin{center}
\subfigure{\includegraphics[width=0.35\textwidth]{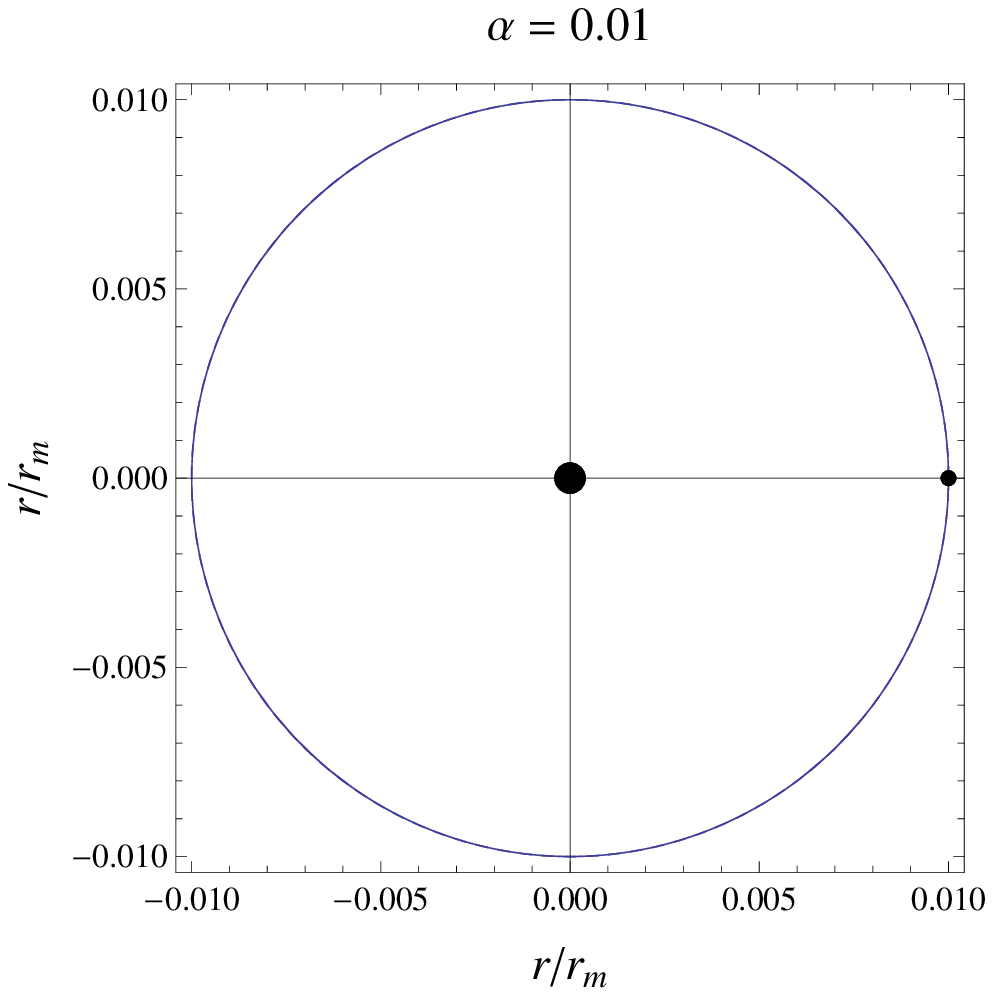}}
\subfigure{\includegraphics[width=0.35\textwidth]{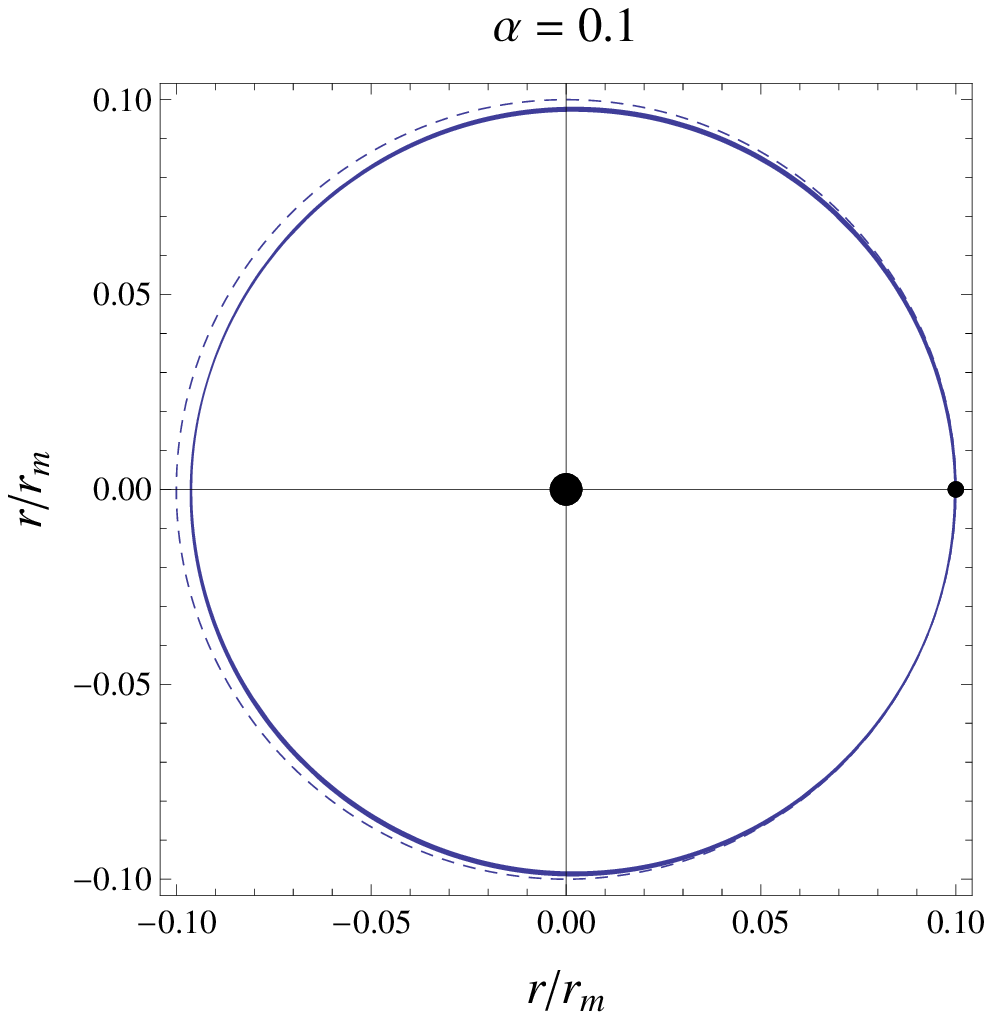}}
\subfigure{\includegraphics[width=0.35\textwidth]{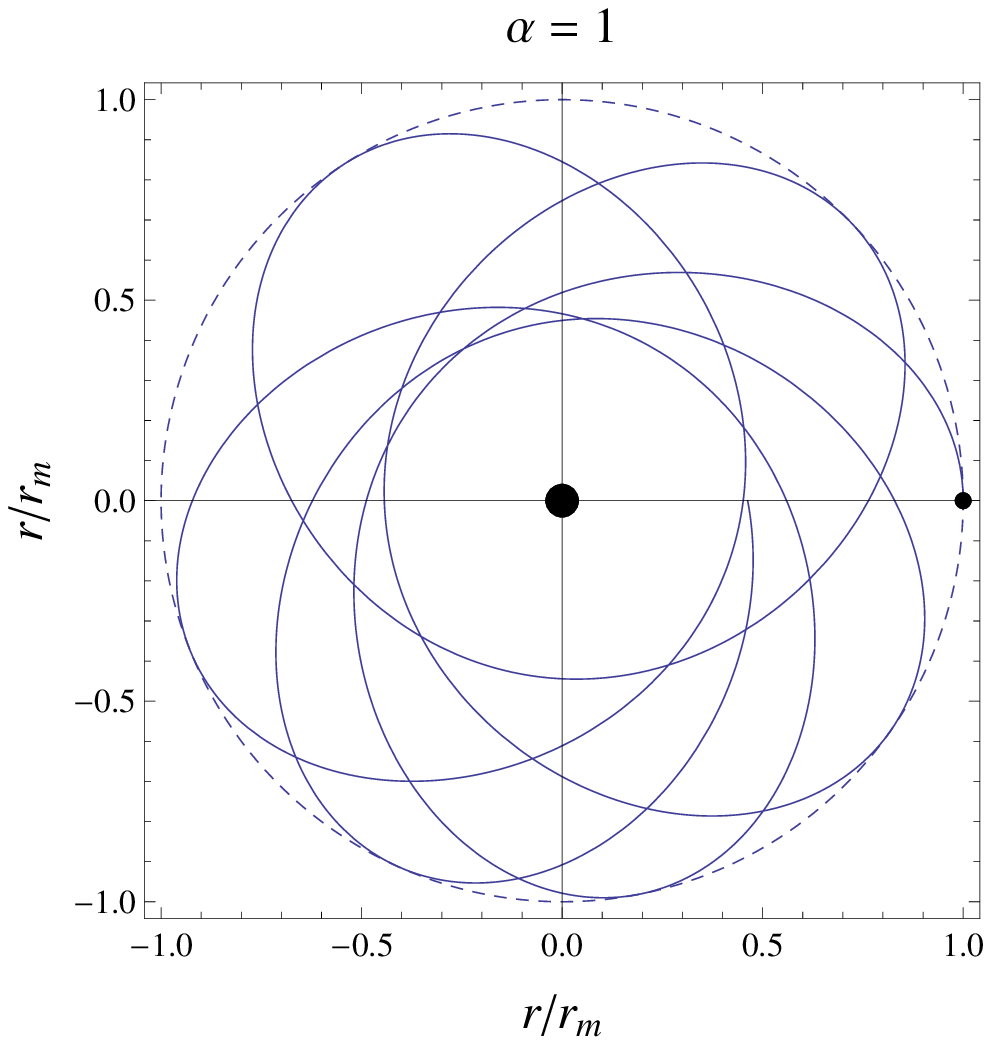}}
\subfigure{\includegraphics[width=0.35\textwidth]{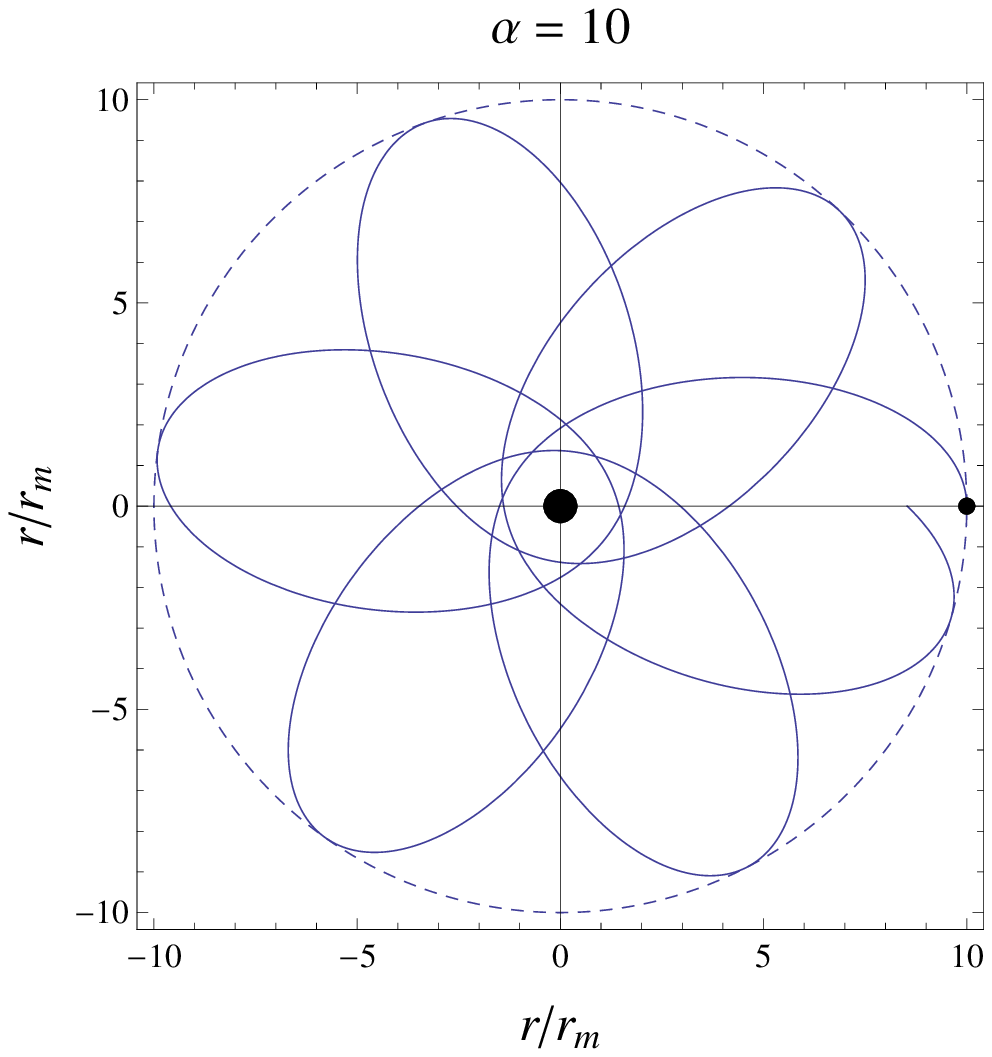}}
\subfigure{\includegraphics[width=0.35\textwidth]{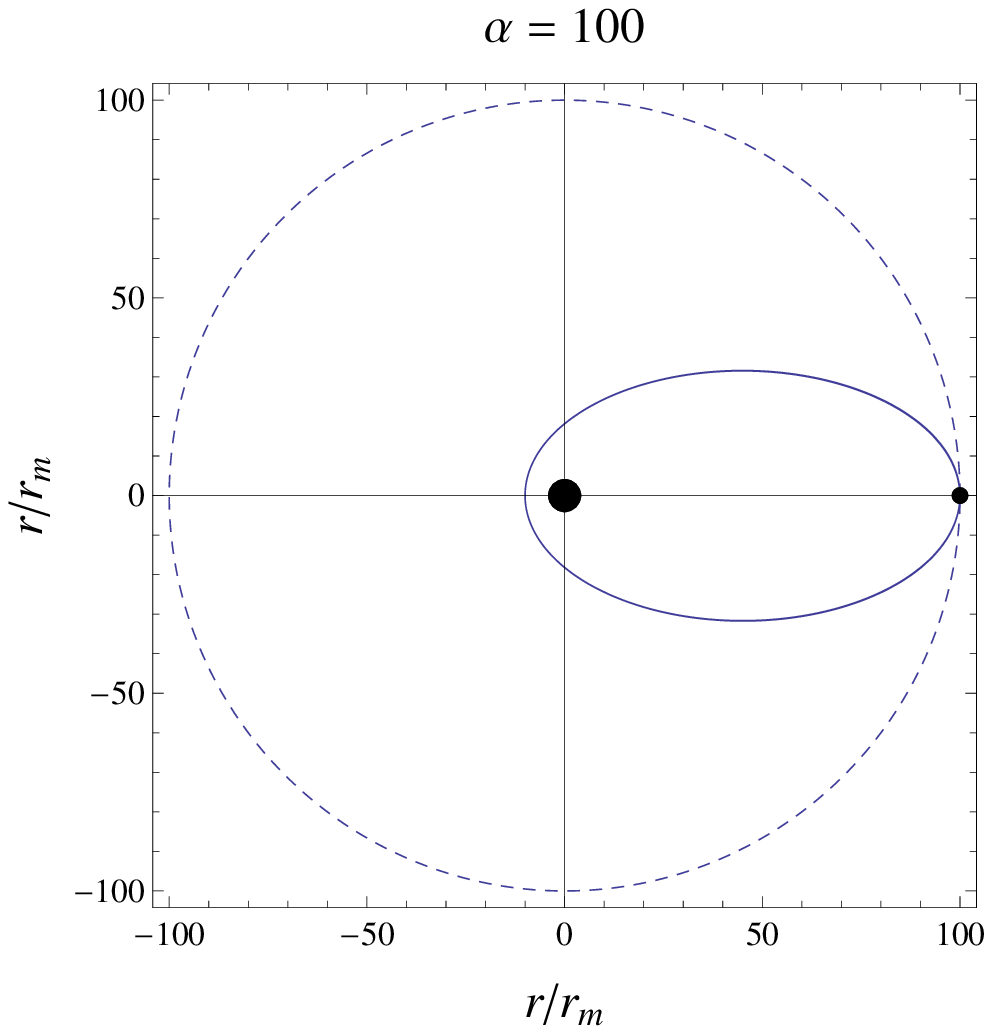}}
\end{center}
\caption{Mirror gravity orbits (solid) for ordinary test body about
  point-like source with $M_{2}/M_{1}=10$, initial radius
  $r_{0}=\alpha r_{m}$ and initial velocity
  $\mathbf{v}_{0}=\sqrt{GM_{1}/r_{0}}{\boldsymbol{\hat\theta}}$.  The
  corresponding Newtonian orbits (dashed) are circular. All orbits are
  shown up to five revolutions.}
\label{mirr_g}
\end{figure*}
The deviation of mirror gravity from an inverse-square law leads to
distinctive orbital dynamics in the gravitational field of a
point-like source, especially at distances $r\sim r_{m}$. Two
prominent features are orbital precession and a varying escape speed
to circular speed ratio, neither of which are present for point-like
sources in classical gravity.

Orbits for an ordinary test body in the gravitational field given by Equations 
(\ref{pot_eq}) and (\ref{accel}) are described by its Binet equation
\begin{equation}\label{binet}
\begin{split}
&\frac{d^{2}u}{d\theta^{2}}+u= \\
& \frac{G}{2h^{2}}\left[M_{1}+M_{2}+\left(M_{1}-M_{2}\right)\left(1+\frac{1}{r_{m}u}\right)e^{-\frac{1}{r_{m}u}}\right],
\end{split}
\end{equation}
where the reciprocal radial distance $u=1/r$ is a function of the
plane polar angle $\theta$, and the specific angular momentum
$h=L/m=r^{2}\dot{\theta}$ is a conserved quantity. Taking the Yukawa
radius $r_{m}=10\;\mathrm{kpc}$ (as assumed by BPR) here and
henceforth, we solve Equation (3) numerically for orbits about a
point-like source with $M_{2}/M_{1}=10$ (see Figure \ref{mirr_g}).

The test body is given an initial radial distance of $r_{0}=\alpha
r_{m}$ and an initial transverse velocity of
$\mathbf{v}_{0}=\sqrt{GM_{1}/r_{0}}{\boldsymbol{\hat\theta}}$, which
is the circular velocity in the absence of mirror matter and mirror
gravity. Its orbit is essentially Newtonian for $\alpha\ll1$, as
gravitational interaction between ordinary and mirror matter is
negligible at small distances, and Equation (\ref{accel}) reduces to
\begin{equation*}
g\left(r\right)=\frac{GM_{1}}{r^{2}}
\end{equation*}
in the limit $r\rightarrow0$.

For $\alpha\gg1$, gravitational interaction between ordinary and
mirror matter becomes universal, and Equation (\ref{accel}) reduces
to
\begin{equation*}
g\left(r\right)=\frac{G\left(M_{1}+M_{2}\right)}{2r^{2}}
\end{equation*}
in the limit $r\rightarrow\infty$. The test body feels the combined
inverse-square field of both matter types (but with an effective
gravitational constant of $G/2$), resulting in an elliptical orbit.

Precessing orbits are manifest for $\alpha\sim0.1$ to $\alpha\sim10$,
and are explicitly non-classical. They may be expected to play a role
in galactic stellar dynamics at distance scales on the order of the
Yukawa radius from the centre of a galaxy.

A more useful signature of mirror gravity, which may allow it to be
observationally corroborated or falsified, lies in the escape speeds
that it predicts. For a point-like source with gravitational field
given by Equations (\ref{pot_eq}) and (\ref{accel}), the escape and
circular speeds are given respectively by
\begin{equation}
\begin{split}
v_{\mathrm{esc}}\left(r\right)&=\sqrt{-2\phi\left(r\right)}\\
&=\sqrt{\frac{G}{r}\left(M_{1}+M_{2}+\left(M_{1}-M_{2}\right)e^{-\frac{r}{r_{m}}}\right)},
\end{split}
\end{equation}
and 
\begin{equation}
\begin{split}
&v_{\mathrm{cir}} \left(r\right)=\sqrt{rg\left(r\right)} \\
&~~=\sqrt{\frac{G}{2r}\left[M_{1}+M_{2}+\left(M_{1}-M_{2}\right)\left(1+\frac{r}{r_{m}}\right)e^{-\frac{r}{r_{m}}}\right]}.
\end{split}
\end{equation}

\begin{figure}
\includegraphics[bb=0bp 0bp 288bp 196bp,scale=0.8]{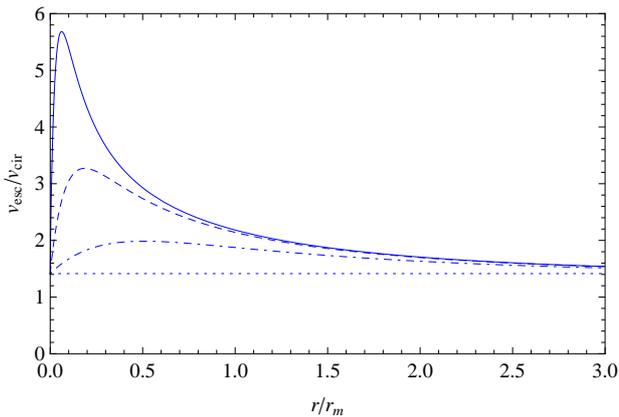}
\label{esc_cir}
\caption{Plot of $v_{\mathrm{esc}}/v_{\mathrm{cir}}$ against $r/r_{m}$ for
$M_{2}/M_{1}=1000$ (solid), $M_{2}/M_{1}=100$ (dashed) and $M_{2}/M_{1}=10$
(dot-dashed). The Newtonian ratio $v_{\mathrm{esc}}/v_{\mathrm{cir}}=\sqrt{2}$
is recovered for $M_{2}/M_{1}=1$ (dotted).}
\end{figure}

In general, the ratio $v_{\mathrm{esc}}/v_{\mathrm{cir}}\neq\sqrt{2}$
and is a function of \emph{r}, in contrast to the Newtonian case. A
plot of $v_{\mathrm{esc}}/v_{\mathrm{cir}}$ for different values of
$M_{2}/M_{1}\geq1$ shows that
$v_{\mathrm{esc}}/v_{\mathrm{cir}}\geq\sqrt{2}$, with the discrepancy
being more pronounced for larger mass ratios (see Figure
\ref{esc_cir}).
\footnote{The cosmological ratio between the dark and visible energy
 densities is $\rho_{2}/\rho_{1}\approx10$ in the
 BPR model. $M_{2}/M_{1}$ should approach the
 cosmological ratio for larger galaxies, while smaller galaxies are
 expected to be more dark matter-dominated. In any case, only values
 of $M_{2}/M_{1}>1$ will be considered in this paper.}  This suggests
that mirror gravity may predict significantly larger escape speeds
compared to classical gravity when both models are fitted to give
similar circular speeds, as is done in the modelling of galactic
rotational curves.

\section{Escape speeds in disc galaxies}\label{esc_speeds_discs}

Mirror matter and mirror gravity have been put forth by BPR as a
collisional and dissipative dark matter candidate capable of
explaining flat galactic rotational curves.  They have successfully
fitted this MDM model to the observed rotational curves of several
disc galaxies, using simplified and similar mass density profiles for
the visible and dark matter distributions. In this section, we obtain
comparable fits using a CDM model of the same complexity, and generate
the corresponding galactic escape speed curves for both models.

The distribution of visible matter in a typical disc galaxy is approximated
most simply as a two-dimensional disc with the exponential surface
mass density
\begin{equation}\label{sigma1}
\Sigma_{1,d}\left(r\right)=\frac{M_{1,d}}{2\pi r_{1,d}^{2}}e^{-\frac{r}{r_{1,d}}},
\end{equation}
where $M_{1,d}$ and $r_{1,d}$ are the total mass and characteristic
length scale of the disc respectively. In the BPR
model, the distribution of mirror dark matter is assumed to take on a
similar profile (i.e. another exponential disc concentric to the
visible disc); this is justifiable since mirror matter has the same
microphysics as ordinary matter and can be expected to undergo the
same process of galaxy formation, up to a rescaling of
parameters. Hence its surface mass density is given
by\begin{equation}\label{sigma2}
\Sigma_{2,d}\left(r\right)=\frac{M_{2,d}}{2\pi
 r_{2,d}^{2}}e^{-\frac{r}{r_{2,d}}},\end{equation} where $M_{2,d}$
and $r_{2,d}$ are the corresponding parameters for the mirror dark
disc.

For an ordinary test body with position vector \textbf{r} in the plane
of the two discs, the acceleration vector follows as
\begin{equation}\label{acc_vec}
\begin{split}
&\mathbf{g}\left(\mathbf{r}\right)=-\frac{G}{2}\int_{\mathrm{disc}}d^{2}\mathbf{r}'\frac{\mathbf{r}-\mathbf{r}'}{\left|\mathbf{r}-\mathbf{r}'\right|^{3}}\left[\vphantom{\frac12}\Sigma_{1,d}\left(r'\right)+\Sigma_{2,d}\left(r'\right)\right.\\
&+\left.\left(\Sigma_{1,d}\left(r'\right)-\Sigma_{2,d}\left(r'\right)\right)\left(1+\frac{\left|\mathbf{r}-\mathbf{r}'\right|}{r_{m}}\right)e^{-\frac{\left|\mathbf{r}-\mathbf{r}'\right|}{r_{m}}}\right],
\end{split}
\end{equation}
where
$\mathbf{g}\left(\mathbf{r}\right)=-g\left(r\right)\hat{\mathbf{r}}$.
The integration can be performed semi-analytically in this case, and
yields the rotational curve $v\left(r\right)=\sqrt{rg\left(r\right)}$
of the model.

\begin{figure*}
\begin{center}
\subfigure{\includegraphics[width=0.47\textwidth]{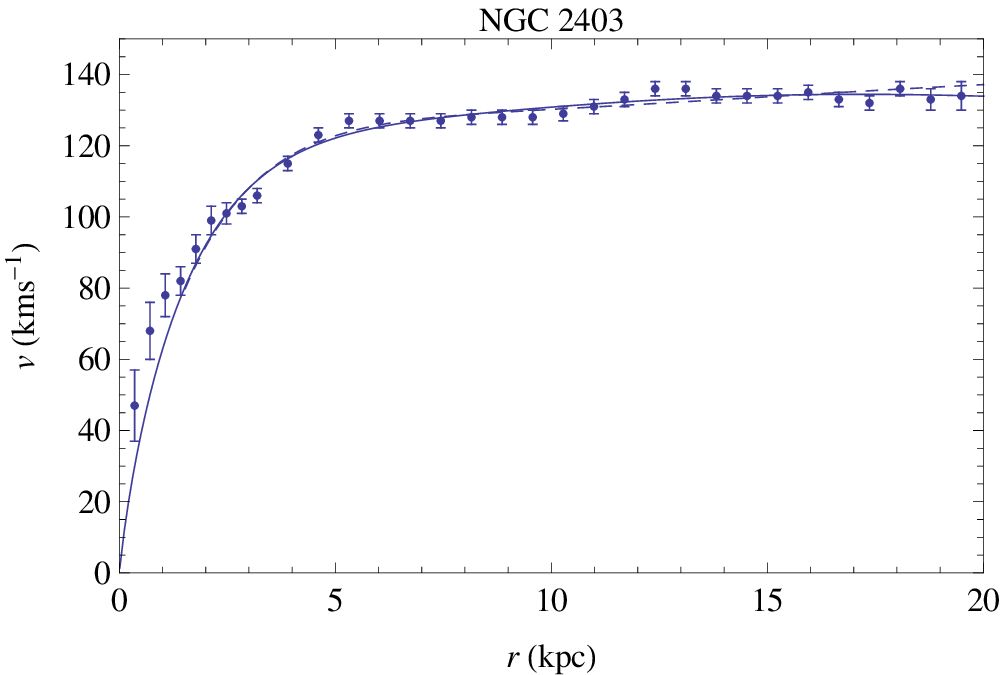}}
\subfigure{\includegraphics[width=0.47\textwidth]{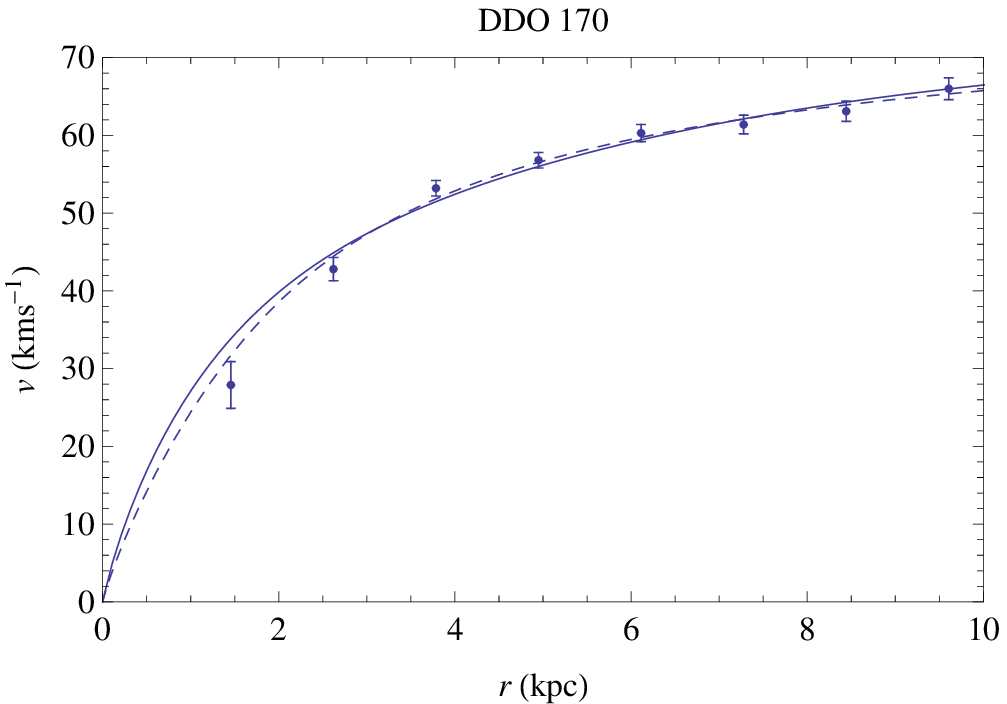}}
\subfigure{\includegraphics[width=0.47\textwidth]{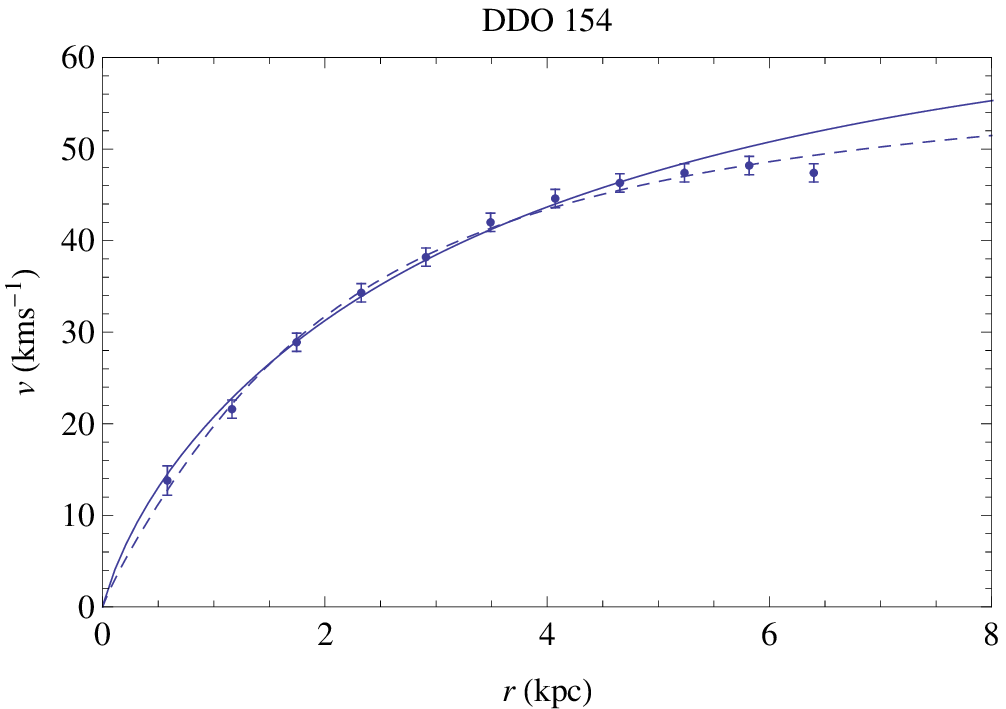}}
\subfigure{\includegraphics[width=0.47\textwidth]{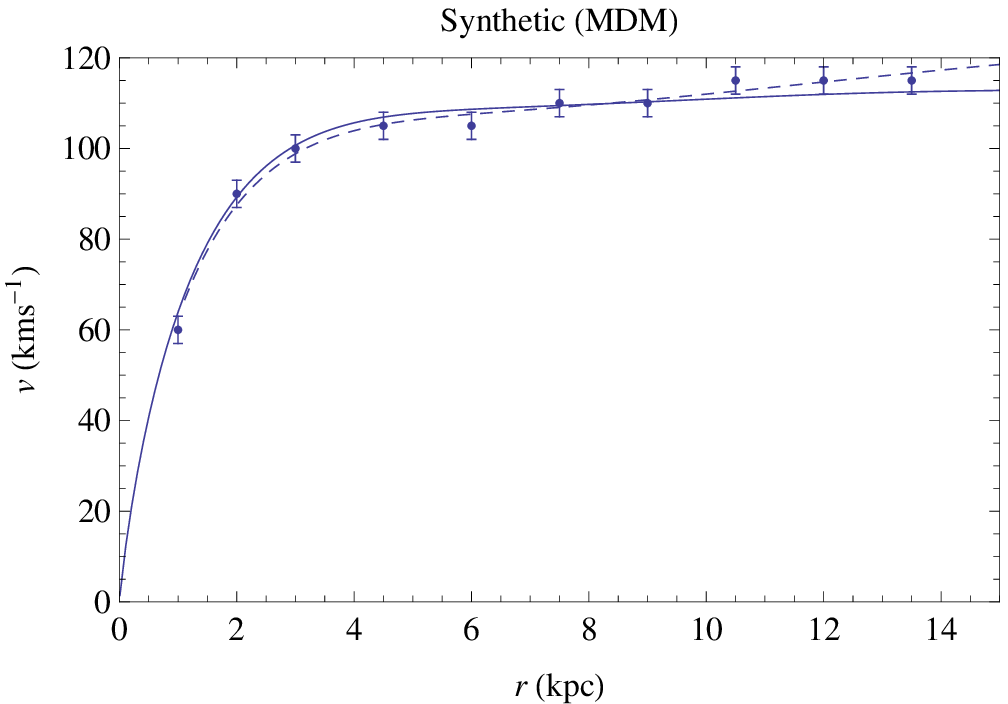}}
\subfigure{\includegraphics[width=0.47\textwidth]{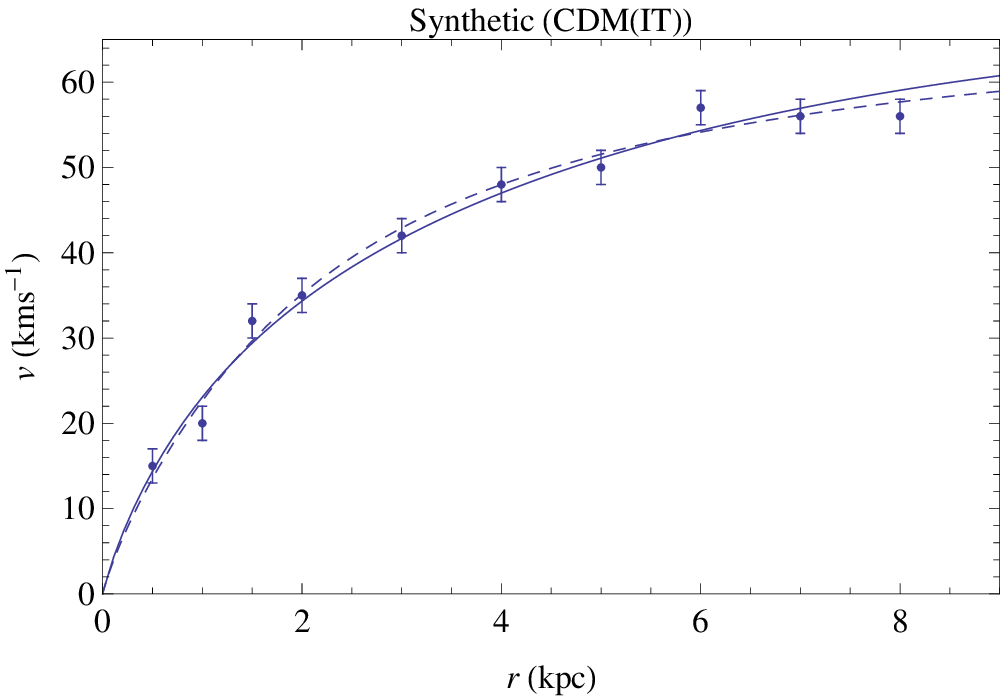}}
\end{center}
\label{rot_curves_disc}
\caption{Observed and synthetic rotational curves for different disc galaxies,
  with both MDM (solid) and CDM(IT) (dashed) three-parameter
  fits. Parameters and values of $\chi_{\mathrm{red}}^{2}$ are given
  in Table \ref{tab_fit}.}
\end{figure*}

Assuming that the mass-to-light ratio $\gamma=M_{1,d}/M_{L}$ of the
visible matter (where the luminous mass $M_{L}$ is an observed fixed
parameter) is constant for each galaxy, the dimensionless quantities
$\gamma$, $\beta=M_{2,d}/M_{1,d}$ and $\alpha=r_{2,d}/r_{1,d}$ have
been taken by BPR as fit parameters for the rotational curves of four
disc galaxies.
\footnote{Rotational curve fits for the large spiral galaxy NGC 7331
  require the contribution of an additional bulge component, and will
  be omitted from the MDM--CDM(IT) comparisons in this section. We
  address the issue of a galactic bulge when modelling the Milky Way
  in Section \ref{esc_MW}.}  The visible length scale $r_{1,d}$ for
each galaxy is deduced from and fixed by observations. The resulting
three-parameter fits are demonstrably good, with
$\chi_{\mathrm{red}}^{2}\approx1$ for each galaxy (see Figure
\ref{rot_curves_disc} and Table \ref{tab_fit}).

Numerous galactic rotational curve fits for CDM models exist and are
well documented in the literature. These are of varying complexity,
but nearly all of them incorporate components such as a galactic bulge
or a gaseous contribution, in addition to the standard galactic disc
and dark halo components. Indeed, the rotational curves of the
galaxies considered by BPR have been fitted for a multi-component CDM
model by \cite{Begeman1991} (henceforth BBS).

We adopt a simplified two-component version of the BBS model in this
paper, allowing direct comparison between the MDM model and a CDM
model of the same complexity. The distribution of visible matter is
again an exponential disc with surface mass density given by Equation
(\ref{sigma1}), while a pseudo-isothermal truncated sphere (which we
shall refer to from now on as CDM(IT)) is assumed for the distribution of
dark matter.

The volume mass density of the dark halo is given by
\begin{equation}\label{dens_halo}
\rho_{h}\left(r\right)=\frac{\rho_{s}}{1+\frac{r^{2}}{r_{s}^{2}}},
\end{equation}
where $\rho_{s}$ and $r_{s}$ are the central density and
characteristic length scale of the halo respectively. This profile
avoids the central cusps inherent in those produced by numerical
\emph{n}-body simulations (e.g. the model by \cite{Navarro1995}), but
has a divergent total mass and must be truncated at a suitable
distance from the galactic centre.

An analytical expression is available for the rotational curve of this
simplified BBS model. We can write
\begin{equation}
v\left(r\right)=\sqrt{v_{\mathrm{disc}}^{2}\left(r\right)+v_{\mathrm{halo}}^{2}\left(r\right)},
\end{equation}
where $v_{\mathrm{disc}}\left(r\right)$ and $v_{\mathrm{halo}}\left(r\right)$
are rotational curves for the visible disc and dark halo respectively.
The contribution from the disc has been obtained by \cite{Freeman1970},
using a method by \cite{Toomre1963}, as
\begin{equation}\label{toomre}
\begin{split}v_{\mathrm{disc}}^{2}\left(r\right)&=\frac{GM_{1,d}r^{2}}{2r_{1,d}^{3}}\left[\vphantom{\frac12}I_{0}\left(\frac{r}{2r_{1,d}}\right)K_{0}\left(\frac{r}{2r_{1,d}}\right)\right.\\
&-\left.I_{1}\left(\frac{r}{2r_{1,d}}\right)K_{1}\left(\frac{r}{2r_{1,d}}\right)\right],
\end{split}
\end{equation}
where the $I_{n}$ and $K_{n}$ are modified Bessel functions evaluated
at $r/2r_{1,d}$. \cite{Begeman1987} gives the contribution from the halo
as
\begin{equation}\label{vhalo}
v_{\mathrm{halo}}^{2}\left(r\right)=v_{\mathrm{max}}^{2}\left(1-\frac{r_{s}}{r}\tan^{-1}\frac{r}{r_{s}}\right),
\end{equation}
where $v_{\mathrm{max}}=\sqrt{4\pi G\rho_{s}r_{s}^{2}}$ is the asymptotic
rotational speed for the halo at $r\gg r_{s}$.

As in the BPR model, we use data from \cite{Begeman1987},
\cite{Carignan1989}, \cite{Lake1990} and \cite{Begeman1991} for the
intermediate spiral galaxy NGC 2403 and the dwarf galaxies DDO 170 and
DDO 154. Again assuming a constant mass-to-light ratio for each galaxy
and taking $M_{L}$ and $r_{1,d}$ as fixed parameters, we fit the
rotational curves of these three disc galaxies by using $\gamma$,
$v_{\mathrm{max}}$ and $r_{s}$ as fit parameters.

Furthermore, we have generated synthetic rotational curves for two
disc galaxies, one whose physics is dictated by MDM and the other by
CDM(IT). Typical observational errors (Gaussian noise) have been
assigned to these synthetic galaxies, which are then treated as actual
observed galaxies and fitted using the procedures outlined in this
section. We find that the MDM fit recovers best-fit values in good
agreement with the actual values for the synthetic MDM galaxy, while
the CDM(IT) fit matches the parameters of the synthetic CDM(IT) galaxy
as well; this verifies the accuracy of our fitting algorithms.
Finally, a side-by-side comparison of all the MDM and CDM(IT) fits
shows that both models are able to reproduce both the observed and
synthetic galactic rotational curves, and equally well (see Figure
\ref{rot_curves_disc} and Table \ref{tab_fit}).

\begin{center}
\begin{table*}

\caption{MDM and CDM(IT) best-fit parameters for the rotational curves
  of different disc galaxies. $M_{L}$ and $r_{1,d}$ are taken as fixed
  parameters, while $\gamma$, $\beta$, $\alpha$, $v_{\mathrm{max}}$
  and $r_{s}$ are taken as fit parameters. Both models provide
  realistic (all values of $\gamma\sim1$) and good (all values of
  $\chi_{\mathrm{red}}^{2}\approx1$) fits. The fitting algorithms
  recover best-fit values in good agreement with the actual values for
  the synthetic disc galaxies (given in parentheses).}

\hspace*{\fill}\begin{tabular}{c c c c c c c}
\hline 
\noalign{\vskip\doublerulesep}
\multicolumn{1}{c}{} &  & NGC 2403 & DDO 170 & DDO 154 & Synthetic (MDM) & Synthetic (CDM(IT))\tabularnewline[\doublerulesep]
\hline 
\noalign{\vskip\doublerulesep}
Fixed parameters & $M_{L}\left(10^{9}M_{\odot}\right)$ & 7.9 & 0.18 & 0.05 & 8.0 & 0.10\tabularnewline[\doublerulesep]
\noalign{\vskip\doublerulesep}
 & $r_{1,d}\left(\mathrm{kpc}\right)$ & 2.0 & 1.1 & 0.5 & 1.6 & 0.8\tabularnewline[\doublerulesep]
\noalign{\vskip\doublerulesep}
MDM fit & $\gamma=M_{1,d}/M_{L}$ & 1.6 & 2.2 & 0.6 & 1.0 (1.0) & 1.2\tabularnewline[\doublerulesep]
\noalign{\vskip\doublerulesep}
 & $M_{1,d}\left(10^{9}M_{\odot}\right)$ & 12.6 & 0.4 & 0.03 & 8.2 (8.0) & 0.12\tabularnewline[\doublerulesep]
\noalign{\vskip\doublerulesep}
 & $\beta=M_{2,d}/M_{1,d}$ & 20.6 & 190 & 2000 & 22.3 (23.0) & 567\tabularnewline[\doublerulesep]
\noalign{\vskip\doublerulesep}
 & $M_{2,d}\left(10^{9}M_{\odot}\right)$ & 260 & 76 & 60 & 183 (184) & 68\tabularnewline[\doublerulesep]
\noalign{\vskip\doublerulesep}
 & $\alpha=r_{2,d}/r_{1,d}$ & 1.6 & 0.8 & 1.3 & 1.6 (1.6) & 1.0\tabularnewline[\doublerulesep]
\noalign{\vskip\doublerulesep}
 & $r_{2,d}\left(\mathrm{kpc}\right)$ & 3.2 & 0.87 & 0.66 & 2.5 (2.6) & 0.8\tabularnewline[\doublerulesep]
\noalign{\vskip\doublerulesep}
 & $\chi_{\mathrm{red}}^{2}$ & 1.3 & 1.5 & 0.6 & 1.3 & 1.3\tabularnewline[\doublerulesep]
\noalign{\vskip\doublerulesep}
CDM(IT) fit & $\gamma=M_{1,d}/M_{L}$ & 1.5 & 1.1 & 0.2 & 1.0 & 0.9 (1.0)\tabularnewline[\doublerulesep]
\noalign{\vskip\doublerulesep}
 & $M_{1,d}\left(10^{9}M_{\odot}\right)$ & 12.1 & 0.2 & 0.01 & 7.7 & 0.09 (0.10)\tabularnewline[\doublerulesep]
\noalign{\vskip\doublerulesep}
 & $v_{\mathrm{max}}\left(\mathrm{kms^{-1}}\right)$ & 155 & 77 & 61 & 129 & 70 (70)\tabularnewline[\doublerulesep]
\noalign{\vskip\doublerulesep}
 & $r_{s}\left(\mathrm{kpc}\right)$ & 5.0 & 2.1 & 1.7 & 4.1 & 2.0 (2.0)\tabularnewline[\doublerulesep]
\noalign{\vskip\doublerulesep}
 & $\chi_{\mathrm{red}}^{2}$ & 1.6 & 1.2 & 0.8 & 1.3 & 1.0\tabularnewline[\doublerulesep]
\hline 
\end{tabular}\hspace*{\fill}

\label{tab_fit}
\end{table*}
\end{center}

Now, for the MDM model, the gravitational potential felt by an ordinary
test body with position vector \textbf{r} in the plane of the two
discs is given by
\begin{equation}\label{phi_MDM}
\begin{split}
\phi_{\mathrm{MDM}}\left(r\right)&=-\frac{G}{2}\int_{\mathrm{disc}}d^{2}\mathbf{r}'\frac{1}{\left|\mathbf{r}-\mathbf{r}'\right|}\big[\Sigma_{1,d}\left(r'\right)+\Sigma_{2,d}\left(r'\right)\\
&+\left.\left(\Sigma_{1,d}\left(r'\right)-\Sigma_{2,d}\left(r'\right)\right)
e^{-\frac{\left|\mathbf{r}-\mathbf{r}'\right|}{r_{m}}}\right],
\end{split}
\end{equation}
where $\phi_{\mathrm{MDM}}\left(r\right)$ depends only on \emph{r}
due to cylindrical symmetry. For the CDM(IT) model, the gravitational
potential felt by a test body with position vector \textbf{r} in the
plane of the disc is given by
\begin{equation}
\phi_{\mathrm{CDM(IT)}}\left(r\right)=-G\int_{\mathrm{disc}}\!\!\!d^{2}\mathbf{r}'\frac{\Sigma_{1,d}\left(r'\right)}{\left|\mathbf{r}-\mathbf{r}'\right|}-G\int_{\mathrm{halo}}\!\!\!d^{3}\mathbf{r}'\frac{\rho_{h}\left(r'\right)}{\left|\mathbf{r}-\mathbf{r}'\right|},
\end{equation}
where $\phi_{\mathrm{CDM(IT)}}\left(r\right)$ depends only on \emph{r}
as well.

\begin{figure*}
\begin{center}
\subfigure{\includegraphics[width=0.44\textwidth]{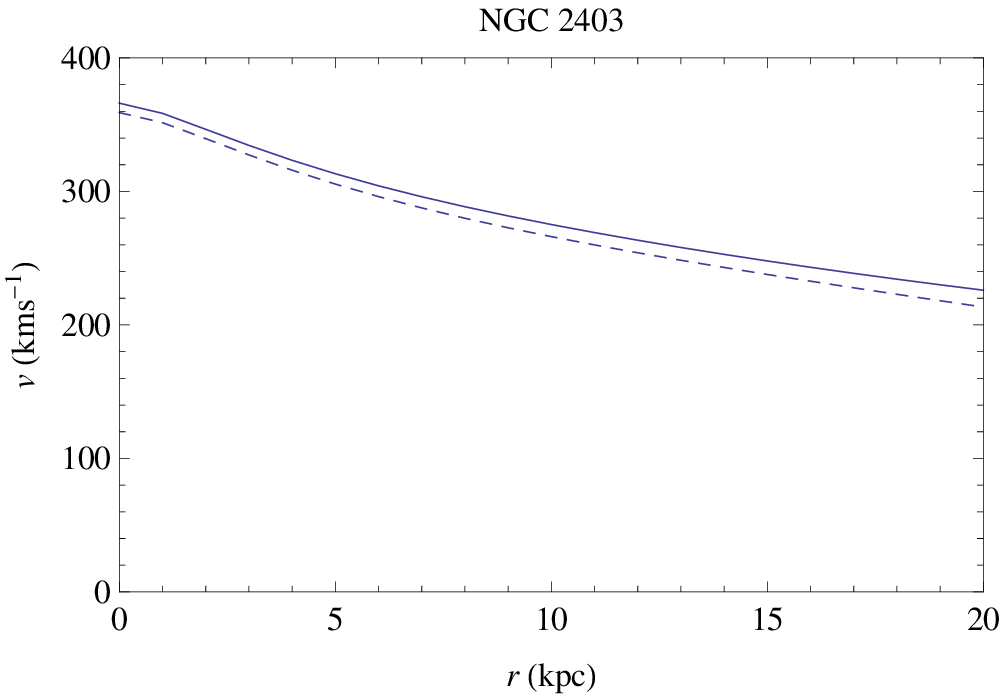}}
\subfigure{\includegraphics[width=0.44\textwidth]{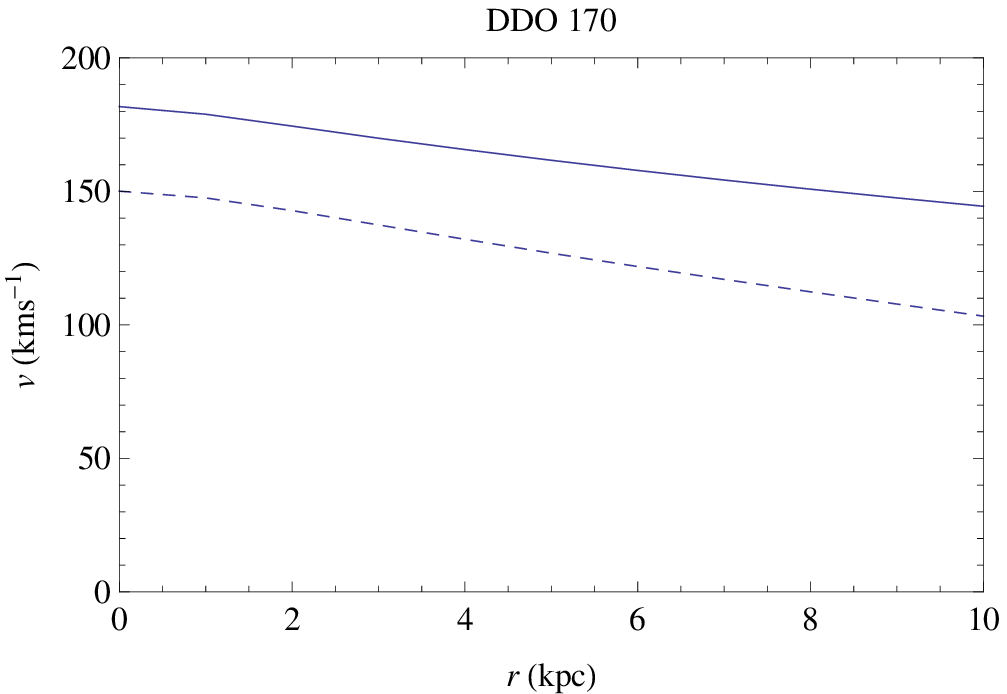}}
\subfigure{\includegraphics[width=0.44\textwidth]{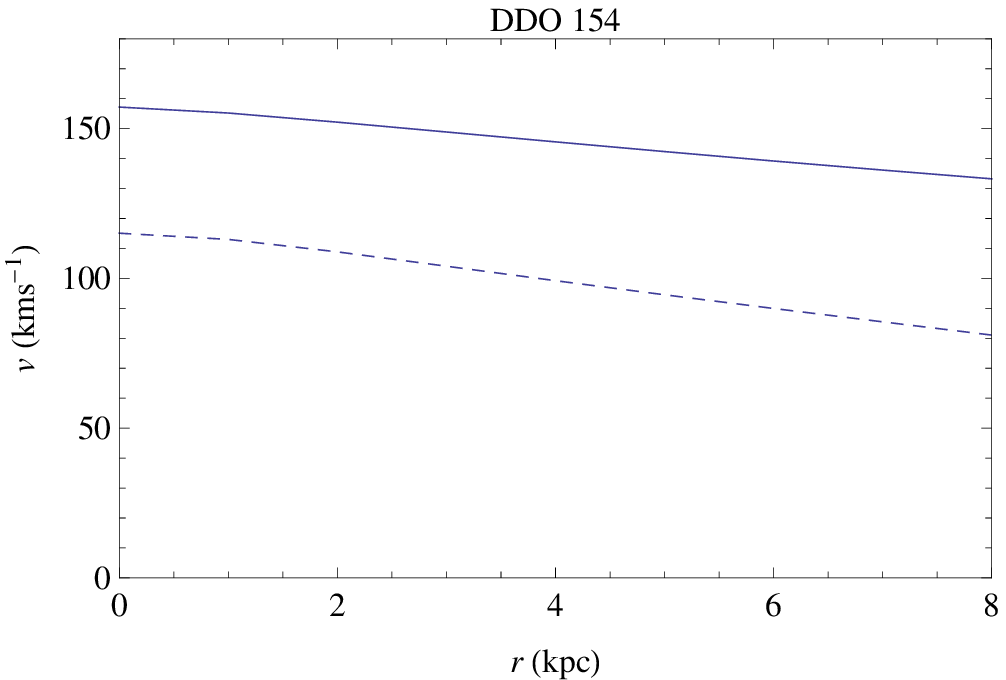}}
\subfigure{\includegraphics[width=0.44\textwidth]{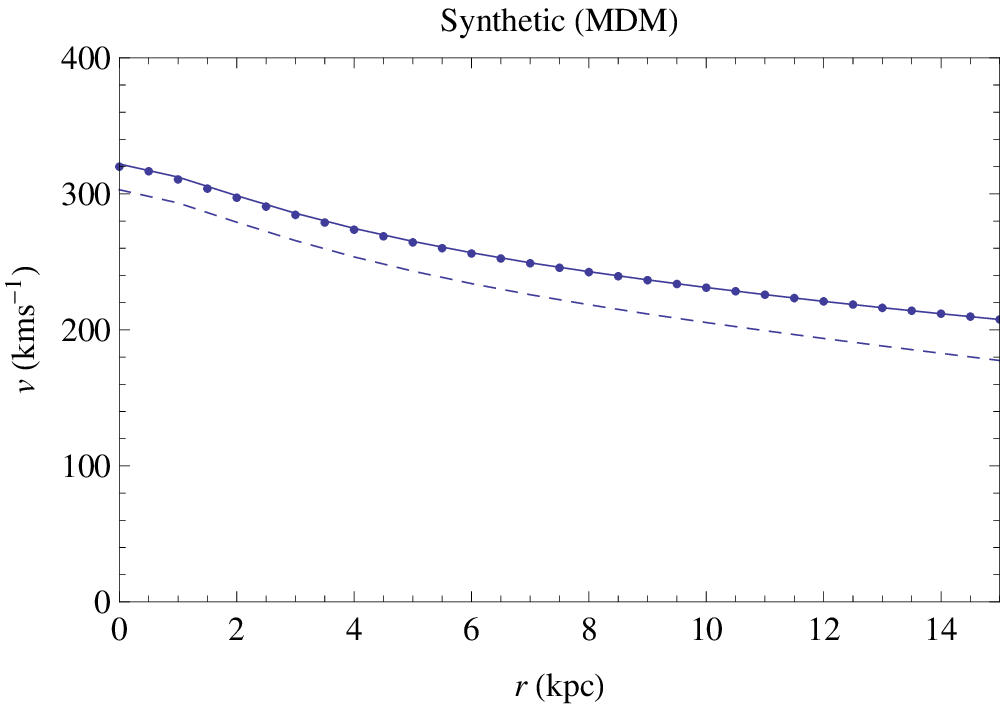}}
\subfigure{\includegraphics[width=0.44\textwidth]{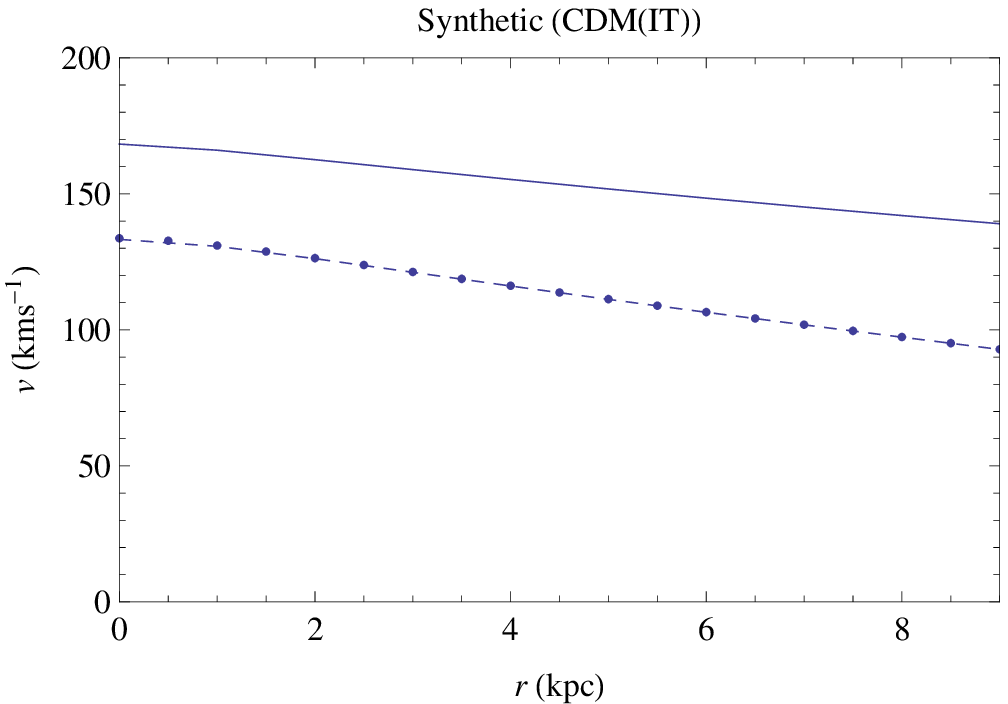}}
\end{center}
\caption{Predicted MDM (solid) and CDM(IT) (dashed) escape speed
  curves for stars in the galactic plane of observed and synthetic
  disc galaxies, using parameters given in \ref{tab_fit}. The actual
  (dotted) escape speed curves for the synthetic galaxies have been
  included as well.}
\label{esc}
\end{figure*}

Using the parameters given in Table \ref{tab_fit}, we obtain MDM and
CDM(IT) escape speed curves $v\left(r\right)=\sqrt{-2\phi\left(r\right)}$
for the five disc galaxies by numerical integration, truncating the
discs and halos at a value of \emph{r} just beyond the farthest
data point for each galaxy (see Figure \ref{esc}).

As anticipated from our discussion of point-like source models in
Section \ref{point_s}, escape speeds for stars in the galactic plane
are mostly higher in the MDM model. The offset between the two models
appears to be nearly constant out to the truncation distance.  It is
both absolutely and relatively larger for the dwarf galaxies ($\Delta
v\geq30\;\mathrm{kms^{-1}}$ and $\Delta
v/v_{\mathrm{CDM(IT)}}\geq0.2$) than for NGC 2403 ($\Delta
v\approx10\;\mathrm{kms^{-1}}$ and $\Delta
v/v_{\mathrm{CDM(IT)}}\approx0.03$).

This systematic escape speed discrepancy may be attributed to the
increasing importance of mirror dark matter over ordinary matter in
the smaller galaxies (as indicated by the higher values of $\beta$),
which in turn is due to typical distance scales in these galaxies
being smaller than the Yukawa scale $r_m$ ($ \sim \rm{10 kpc}$) such
that dark matter is less effective in gravitationally binding to
ordinary matter.  In the largest galaxies such as the Milky Way (see Section 4), the ratio of the
energy density of mirror dark matter to visible matter
$\rho_{2}/\rho_{1}$ approaches the cosmological value $\sim 10$ as
required by the BPR model.

Crucially, we have also observed in our analysis that escape speeds
for the MDM model do not depend significantly on the choice of
truncation radius, and are constrained by the rotational curve
fits. Thus, observations of the stellar escape speeds in disc galaxies
of different sizes could prove key in establishing---or
abolishing---the validity of the MDM model.

\section{Escape speeds in the Milky Way}\label{esc_MW}

Although the rotational curves of numerous external disc galaxies have
been obtained to a good degree of precision, there are no direct
estimates of their escape speeds at the present time. Conversely,
while the rotational curve of our own galaxy is less established,
efforts have been made by \cite{Smith2007} to constrain the escape
speed at the solar neighbourhood through the Radial Velocity
Experiment (RAVE) survey. The current estimate is significantly in
excess of $\sqrt{2}v_{\mathrm{cir}}$ (where
$v_{\mathrm{cir}}\approx220\;\mathrm{kms^{-1}}$ is the local circular
speed), and has been taken as independent evidence in favour of CDM
models. In this section, we repeat the MDM--CDM(IT) comparison for the
Milky Way, and show that the MDM model is also able to predict a local
escape speed comparable to the currently observed value.

As in Section \ref{esc_speeds_discs}, the visible disc of the Galaxy in
both models is taken as an exponential disc and described by Equation
(\ref{sigma1}). Likewise, the density profiles of the mirror dark disc in the
MDM model and the dark halo in the CDM(IT) model are given respectively by
Equations (\ref{sigma2}) and (\ref{dens_halo}). In both models,
however, an additional bulge component is required to reproduce the
observed Galactic rotational curve.

We employ the tractable Plummer sphere for this purpose, in lieu of
the truncated power-law bulges used in more detailed mass models.
Hence the gravitational potential of the visible bulge is given by

\begin{equation}\label{bulge}
\phi_{\mathrm{bulge}}\left(r\right)=-\frac{GM_{1,b}}{\sqrt{r^{2}+r_{1,b}^{2}}},
\end{equation}
where $M_{1,b}$ and $r_{1,b}$ are the total mass and characteristic
length scale of the bulge respectively. Its volume mass density follows
as
\begin{equation}\label{dens1_bulge}
\rho_{1,b}\left(r\right)=\frac{3M_{1,b}}{4\pi r_{1,b}^{3}}\left(1+\frac{r^{2}}{r_{1,b}^{2}}\right)^{-\frac{5}{2}}
\end{equation}
from Poisson's equation.

By assumption, the presence of a visible bulge necessitates the addition
of a mirror dark bulge component to the MDM model. This is similarly described
by
\begin{equation}\label{dens2_bulge}
\rho_{2,b}\left(r\right)=\frac{3M_{2,b}}{4\pi r_{2,b}^{3}}\left(1+\frac{r^{2}}{r_{2,b}^{2}}\right)^{-\frac{5}{2}},
\end{equation}
where $M_{2,b}$ and $r_{2,b}$ are the corresponding parameters for
the dark bulge. The acceleration vector of an ordinary test body with
position vector \textbf{r} in the Galactic plane follows as
\begin{equation}\label{a_vec}
\mathbf{g}\left(\mathbf{r}\right)=\mathbf{g}_{\mathrm{disc}}\left(\mathbf{r}\right)+\mathbf{g}_{\mathrm{bulge}}\left(\mathbf{r}\right),
\end{equation}
where $\mathbf{g}_{\mathrm{disc}}\left(\mathbf{r}\right)$ is given by
Equation (\ref{acc_vec}) and
$\mathbf{g}_{\mathrm{bulge}}\left(\mathbf{r}\right)$ is obtained in
identical fashion by numerically integrating Equations
(\ref{dens1_bulge}) and (\ref{dens2_bulge}) over the bulge.

With the addition of the visible bulge, the rotational curve of the
CDM(IT) model becomes
\begin{equation}\label{CDM(IT)_rot}
v\left(r\right)=\sqrt{v_{\mathrm{disc}}^{2}\left(r\right)+v_{\mathrm{halo}}^{2}\left(r\right)+v_{\mathrm{bulge}}^{2}\left(r\right)},
\end{equation}
where $v_{\mathrm{disc}}^{2}\left(r\right)$ and
$v_{\mathrm{halo}}^{2}\left(r\right)$ are given by Equations
(\ref{toomre}) and (\ref{vhalo}) as before, and
$v_{\mathrm{bulge}}^{2}\left(r\right)$ follows from Equation
(\ref{bulge}) as
\begin{equation}\label{v_bulge}
v_{\mathrm{bulge}}^{2}\left(r\right)=r\frac{d}{dr}\phi_{\mathrm{bulge}}\left(r\right)=\frac{GM_{1,b}r^{2}}{r_{1,b}^{3}}\left(1+\frac{r^{2}}{r_{1,b}^{2}}\right)^{-\frac{3}{2}}.
\end{equation}

For the MDM model, we assume a constant dark-to-visible mass ratio
(such that $\beta=M_{2,i}/M_{1,i}$) and a constant dark-to-visible
length scale ratio (such that $\alpha=r_{2,i}/r_{1,i}$) in the Galaxy.
We use $M_{1,d}$, $M_{1,b}$, $\beta$ and $\alpha$ as fit parameters,
and take $r_{1,d}=3\;\mathrm{kpc}$ and $r_{1,b}=0.3\;\mathrm{kpc}$ in
rough accordance with observation and other mass models
(e.g. \cite{McMillan2011}).

\begin{figure}
\includegraphics[bb=0bp 0bp 295bp 196bp,scale=0.8]{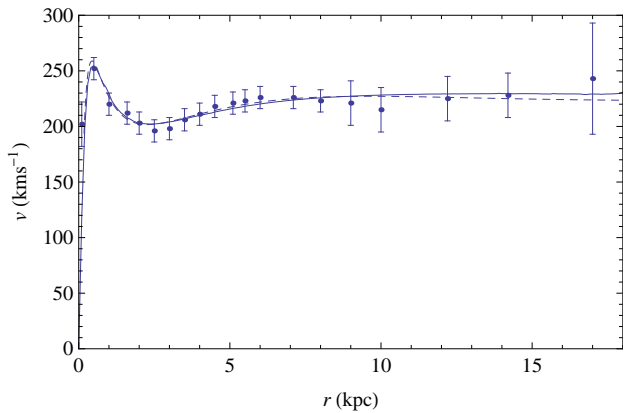}
\caption{Observed Galactic rotational curve out to $2R_{\odot}$, with
both MDM (solid) and CDM(IT) (dashed) four-parameter fits. Parameters
and values of $\chi_{\mathrm{red}}^{2}$ are given in Table
2.}
\label{G_rot}
\end{figure}

For the CDM(IT) model, we use $M_{1,d}$, $M_{1,b}$, $v_{\mathrm{max}}$
and $r_{s}$ as fit parameters, and take $r_{1,d}$ and $r_{1,b}$ as
fixed parameters with the MDM values. With data out to $2R_{\odot}$
(where $R_{\odot}=8.5\;\mathrm{kpc}$) from \cite{Burton1978},
\cite{Blitz1982} and \cite{Clemens1985}, we are able to generate good
rotational curve fits for both models (see Figure \ref{G_rot} and
Table \ref{MDM_CDM}).

\begin{center}
\begin{table}

\caption{MDM and CDM(IT) best-fit parameters for the Galactic rotational curve
out to $2R_{\odot}$. $r_{1,d}$ and $r_{1,b}$ are taken as fixed
parameters, while $M_{1,d}$, $M_{1,b}$, $\beta$, $\alpha$, $v_{\mathrm{max}}$
and $r_{s}$ are taken as fit parameters. Both models provide good
($\chi_{\mathrm{red}}^{2}\approx1$) fits that agree with observation
and other mass models.}
\hspace*{\fill}\begin{tabular}{c c c}
\hline 
& Parameters & Values \\
\hline
\noalign{\vskip\doublerulesep}
Fixed parameters & $r_{1,d}\left(\mathrm{kpc}\right)$ & 3\tabularnewline[\doublerulesep]
\noalign{\vskip\doublerulesep}
& $r_{1,b}\left(\mathrm{kpc}\right)$ & 0.3\tabularnewline[\doublerulesep]
\noalign{\vskip\doublerulesep}
MDM fit & $M_{1,d}\left(10^{9}M_{\odot}\right)$ & 48\tabularnewline[\doublerulesep]
\noalign{\vskip\doublerulesep}
& $M_{1,b}\left(10^{9}M_{\odot}\right)$ & 13\tabularnewline[\doublerulesep]
\noalign{\vskip\doublerulesep}
& $\beta=M_{2,i}/M_{1,i}$ & 12\tabularnewline[\doublerulesep]
\noalign{\vskip\doublerulesep}
& $M_{2,d}\left(10^{9}M_{\odot}\right)$ & 576\tabularnewline[\doublerulesep]
\noalign{\vskip\doublerulesep}
& $M_{2,b}\left(10^{9}M_{\odot}\right)$ & 156\tabularnewline[\doublerulesep]
\noalign{\vskip\doublerulesep}
& $\alpha=r_{2,i}/r_{1,i}$ & 1.7\tabularnewline[\doublerulesep]
\noalign{\vskip\doublerulesep}
& $r_{2,d}\left(\mathrm{kpc}\right)$ & 5.1\tabularnewline[\doublerulesep]
\noalign{\vskip\doublerulesep}
& $r_{2,b}\left(\mathrm{kpc}\right)$ & 0.51\tabularnewline[\doublerulesep]
\noalign{\vskip\doublerulesep}
& $\chi_{\mathrm{red}}^{2}$ & 1.3\tabularnewline[\doublerulesep]
\noalign{\vskip\doublerulesep}
CDM(IT) fit & $M_{1,d}\left(10^{9}M_{\odot}\right)$ & 54\tabularnewline[\doublerulesep]
\noalign{\vskip\doublerulesep}
& $M_{1,b}\left(10^{9}M_{\odot}\right)$ & 11.8\tabularnewline[\doublerulesep]
\noalign{\vskip\doublerulesep}
& $v_{\mathrm{max}}\left(\mathrm{kms^{-1}}\right)$ & 241\tabularnewline[\doublerulesep]
\noalign{\vskip\doublerulesep}
& $r_{s}\left(\mathrm{kpc}\right)$ & 6.6\tabularnewline[\doublerulesep]
\noalign{\vskip\doublerulesep}
& $\chi_{\mathrm{red}}^{2}$ & 1.1\tabularnewline[\doublerulesep]
\hline
\end{tabular}\hspace*{\fill}

\label{MDM_CDM}
\end{table}
\end{center}


Although the fits are simplistic, they are realistic enough to serve
as adequate models of the Galaxy. The total visible (stellar) mass for
the MDM fit is $6.1\times10^{10}\; M_{\odot}$ with a bulge
contribution of $1.3\times10^{10}\; M_{\odot}$; these values are
comparable to estimates by \cite{Gerhard2002}, \cite{Flynn2006},
\cite{Widrow2008} and \cite{McMillan2011}.  The dark-to-visible mass
ratio is $\beta=12$, which approaches the cosmological ratio
$\rho_{2}/\rho_{1}\approx10$ in support of the hypothesis by BPR. The
local surface mass density (at $R_{\odot}=8.5\;\mathrm{kpc}$) of the
disc is $50\; M_{\odot}\mathrm{pc}^{-2}$, in good agreement with
estimates by \cite{Kuijken1989}, \cite{Flynn1994}, \cite{Bienayme2006}
and \cite{Flynn2006}.

The total visible mass for the CDM(IT) fit is $6.6\times10^{10}\;
M_{\odot}$ with a bulge contribution of $1.2\times 10^{10}\;
M_{\odot}$, while the local surface mass density of the disc is $56\;
M_{\odot}$ $\mathrm{pc}^{-2}$.  Again, these values match up reasonably
well with other estimates.  The local dark matter density is
$0.35\;\mathrm{GeVcm}^{-3}$, which is towards the low end of bounds
given by \cite{Catena2010} and \cite{McMillan2011}.

\begin{figure}
\includegraphics[bb=0bp 0bp 295bp 196bp,scale=0.8]{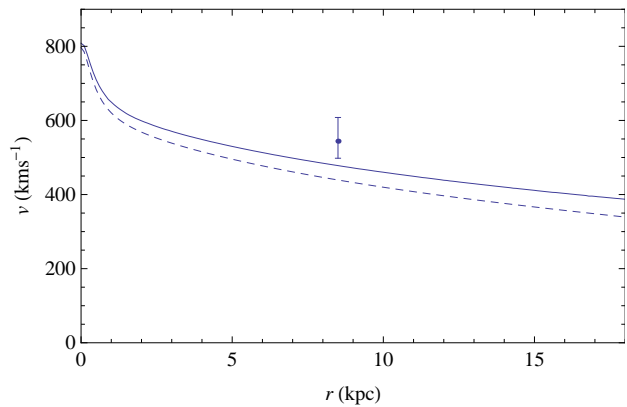}
\caption{Predicted MDM (solid) and CDM(IT) (dashed) escape speed curves out to
$2R_{\odot}$ for stars in the Galactic plane, using the parameters given
in Tab. 2. The RAVE value $v_{\mathrm{esc}}=544_{-46}^{+64}\;\mathrm{kms^{-1}}$
for the local escape speed (at $R_{\odot}=8.5\;\mathrm{kpc}$) is
included as a reference point.}
\label{MDM_CDM_esc}
\end{figure}

Using the parameters given in Table \ref{MDM_CDM}, we obtain MDM and
CDM(IT) escape speed curves for the Galaxy as in Section
\ref{esc_speeds_discs}, but with a bulge term added to the
gravitational potential (see Figure \ref{MDM_CDM_esc}). For the MDM
model, $\phi_{\mathrm{bulge}}\left(r\right)$ is obtained in identical
fashion to Equation (\ref{phi_MDM}) by numerically integrating
Equations (\ref{dens1_bulge}) and (\ref{dens2_bulge}) over the bulge;
for the CDM(IT) model, it is given by Equation (\ref{bulge}).

Again, escape speeds for stars in the Galactic plane are higher in the
MDM model, with an offset that grows from
$\approx20\;\mathrm{kms^{-1}}$ to $\approx50\;\mathrm{kms^{-1}}$ but
remains nearly constant beyond $r=R_{\odot}$. The relative offset at
$r=R_{\odot}$ is $\Delta v/v_{\mathrm{CDM(IT)}}\approx0.09$.  Both models
predict values for the local escape speed that fall just short of the
lower bound on the RAVE value
$v_{\mathrm{esc}}=544_{-46}^{+64}\;\mathrm{kms^{-1}}$; in the CDM(IT)
case, the fairly large discrepancy is due to the truncation
of the dark halo at $r=2R_{\odot}$. Indeed, by extending the truncation radius of the halo
 to 45 kpc, we find an escape speed for the CDM(IT) model that matches the 
 observed RAVE value. Regardless, the fact that the MDM
model is able to predict a local value of $v_{\mathrm{esc}}$
significantly greater than $\sqrt{2}v_{\mathrm{cir}}$ corroborates the
RAVE measurement, and lends credence to its own viability as a dark
matter model.

\section{Discussion and conclusions}

The addition of a non-luminous component of matter distributed
spherically about the centre of a galaxy and extending well beyond the
luminous disc component (i.e. a dark matter halo) explained the flat
rotational curves of disc galaxies in the framework of Newtonian
gravity, and led to the CDM paradigm in cosmology.  In these models, a
significant fraction of the gravitating matter in the Universe is in a
dark collisionless form, with a density parameter ratio between dark
and baryonic matter of $\Omega_d/\Omega_b \sim
5$. The CDM model has been extremely successful in explaining early
Universe thermodynamics, through to fluctuations in the CMB and the
formation of structures on cosmological scales. Any viable alternative
model must not only explain galactic rotational curves, but also match
the successes of CDM in large-scale cosmology.

A much-discussed alternative for explaining galaxy rotational curves
is the modified Newtonian dynamics (MOND) model of \cite{Milgrom1983},
which entertains the possibility of an acceleration due to gravity
that departs from the Newtonian form in regimes where the dynamical
accelerations are small. An in-depth review of the theory is provided
by \cite{Sanders2002}. First proposed as a phenomenological model, the
essence of MOND has subsequently been incorporated in a theory of
modified gravity (MOG) by \cite{Moffat2005}, which is based on a
covariant generalisation of Einstein's theory with the addition of
auxiliary gravitational fields to the metric. MOG has been
successfully used to model a large sample of galactic rotational
curves \cite{Brownstein2006a} and galaxy clusters
\cite{Brownstein2006b}. However, these models that focus on the
gravitational force appear not to be as well-placed to address
questions on early Universe thermodynamics and large-scale cosmology.

On the other hand, the MDM model of BPR is based on
a fundamental theory of all particles and forces; as with the CDM
model, it appears to have the potential to not only explain galaxy
rotational curves, but also to address problems relating to the early
history of the Universe.  In proposing their model,
BPR point out that to date there has been no
observational evidence for the dark matter postulated in CDM
models. They note that one of the major unanswered questions in CDM
cosmology, namely the reason for the closeness of the observed ratio
$\Omega_d/\Omega_b$ to unity, may find an easier solution in MDM
models---where the dark (mirror) and baryonic (ordinary) sectors are
expected to be closely linked. They also note that the density
profiles predicted by \emph{n}-body simulations for the CDM halos in
galaxies show a central cusp, for which there is no observational
evidence from rotational curves of low surface brightness galaxies.
While acknowledging that this may be a problem with the numerical
simulations of galaxy formation in the CDM scenario, they point out
that dark matter is collisional in the MDM model, and such cusps are
not expected.

Rotational curves, being based on the circular speed (usually of gas and sometimes augmented
by that of stars) in galaxies, cannot by themselves provide strong constraints on dark
matter models or gravity models. Indeed, CDM, MDM and MOG models have
all been successfully used to model rotational curves of both low and
high surface brightness galaxies. In this paper, we have sought to
highlight some of the differences that may be expected in the dynamics
of stars in galactic discs, which may potentially be used to
discriminate between the CDM and MDM models. In particular, we have focused
on stellar escape speeds from the galactic disc. Following
BPR, our basic premise is that due to the
collisional and dissipative nature of MDM, matter in the two sectors
will have similar density distributions up to a scaling factor. A consequence
is that the MDM model has a very definite prediction on escape speeds that
is readily falsifiable. Notwithstanding the uncertainties inherent in modelling 
the structure and extent of the CDM halos, we predict systematic and measurable 
differences in escape speed between the two models.

We have taken the galaxies NGC 2403, DDO 170 and DDO 154 (whose
rotational curves have been modelled by BPR using MDM) along with a
couple of synthetic galaxies as representative of spiral galaxies with
a range of sizes, and used them to calculate and compare stellar
escape speeds between the MDM and CDM(IT) models. We find that the
escape speeds from the galactic disc are typically higher in the MDM
model, with the percentage increase in escape speed relative to what
is expected from an equivalent CDM(IT) model approaching $\Delta
v/v_{\mathrm{CDM}}\geq 20$ percent in the smaller low surface
brightness galaxies. A similar procedure has been adopted for the
Milky Way, where in addition to the MDM stellar disc, we have also
included an MDM bulge component.

In the case of the Milky Way, recent estimates from the RAVE survey
yield an escape speed from the solar neighbourhood of
$v_{\mathrm{esc}}=544_{-46}^{+64}\;\mathrm{kms^{-1}}$, and an escape
to circular speed ratio $v_{\mathrm{esc}}/v_{\mathrm{cir}}\sim 2.5 $.
If one assumes normal gravity and axially symmetric density
structures, the circular speed gives a measure of the amount of matter
within the solar circle, while the escape speed gives a measure of the
amount of matter outside the solar circle. A measurement of
$v_{\mathrm{esc}}/v_{\mathrm{cir}}$ that is significantly larger than
$\sqrt 2$ could then be taken as independent evidence that there is a
large proportion of mass outside the solar circle. However, we have shown 
that our MDM model of the Milky Way yields a similar value
$v_{\mathrm{esc}}/v_{\mathrm{cir}}\sim 2.2 $, and that given the errors, this
particular test cannot be used to establish the presence of a dark
halo or to distinguish between the two models at the present
time.

This situation is expected to change when the Gaia--ESO public
spectroscopic survey using FLAMES becomes available in 2016
\cite{Gilmore2012}; the survey will yield accurate position and
velocity information of a large number of individual stars in the thin
and thick discs, the bulge and the halo. Detailed studies of the
motions of stars in the solar neighbourhood (as has been done with
RAVE) will allow a more precise measurement of the local escape
speed. With the new set of data, the study could be extended to
establish the radial dependence of the escape velocity, which will
provide even stronger constraints on the different dark matter
models. We may expect substantial differences under the MDM model in
the kinematics of stars in the galactic halo, which may also be
measurable in future surveys.
 
The measurement of escape speeds from discs in external galaxies may
also soon become possible. This will require the accumulation of
position and velocity data for a large number of individual stars
using multi-object optical and infrared spectroscopy with large
optical telescopes, along the lines of the recent spectroscopic study
of the nearby dwarf irregular galaxy WLM using the VLT and Keck II
\cite{Leaman2012}.  Statistical studies of the resulting stellar
velocity distributions of a number of such galaxies viewed at
different inclinations could then be used to estimate escape speeds
and to test the MDM paradigm.

\begin{acknowledgements}
We wish to thank Ken Freeman for helpful comments and discussion. We
also wish to thank the anonymous Referee for his/her constructive
suggestions which have greatly improved the manuscript.
\end{acknowledgements}



\end{document}